\documentclass[%
 reprint,
 amsmath,amssymb,
 prm,
]{revtex4-1}

\usepackage{graphicx}% Include figure files
\usepackage{dcolumn}% Align table columns on decimal point
\usepackage{bm}% bold math
\begin{document}

\preprint{APS/123-QED}

\title{Blue-Green Emission from Epitaxial Yet Cation-Disordered ZnGeN$_{2-x}$O$_{x}$}% Force line breaks with \\

\author{C. L. Melamed\textsuperscript{\textit{1,2}}}
    \email{cmelamed@mymail.mines.edu}
\author{M. B. Tellekamp\textsuperscript{\textit{2}}}
\author{J. S. Mangum\textsuperscript{\textit{1}}}%
\author{J. D. Perkins\textsuperscript{\textit{2}}}
\author{P. Dippo\textsuperscript{\textit{2}}}
\author{E. S. Toberer\textsuperscript{\textit{1,2}}}
\author{A. C. Tamboli\textsuperscript{\textit{2,1}}}

\affiliation{%
 \textsuperscript{1}\,Physics Department, Colorado School of Mines\\
 \textsuperscript{2}\,National Renewable Energy Laboratory
}%

\date{\today}% It is always \today, today,
             %  but any date may be explicitly specified

\begin{abstract}

ZnGeN$_2$ offers a low-cost alternative to InGaN with the potential for bandgap tuning to span the green gap using cation site ordering. The addition of oxygen on the anion site creates an additional degree of electronic tunability. Here, we investigate the structure and optoelectronic properties of an epitaxial ZnGeN$_{2-x}$O$_{x}$ thin film library grown by combinatorial co-sputtering on c-Al$_2$O$_3$. Samples exhibit X-ray diffraction patterns and X-ray pole figures characteristic of a wurtzite (cation-disordered) structure with the expected 6-fold in-plane symmetry. Transmission electron microscopy reveals a semi-coherent interface with periodic dislocations that relieve strain from the large lattice mismatch, and confirms the in-plane and out-of-plane crystallographic orientation. Room-temperature photoluminescence exhibits peaks between 2.4 and 2.8\,eV which are consistent with a sharp absorption onset observed by UV-vis spectroscopy. These results demonstrate low-cost synthesis of optically active yet cation disordered ZnGeN$_{2-x}$O$_{x}$, indicating a path toward application as a blue-green emitter.

\end{abstract}

%\keywords{Suggested keywords}%Use showkeys class option if keyword
                              %display desired
\maketitle

%\tableofcontents

\section{\label{sec:intro}Introduction}

The III-V family of semiconductors revolutionized optoelectronics, from high-efficiency GaAs solar cells to the GaN-based LEDs which won the 2014 Nobel Prize for their record-breaking brightness and energy efficiency.\cite{Essig2017,Nobel2014} However, there are serious limitations to the III-Vs, including (1) the requirement for low defect density which necessitates expensive epitaxial growth, and (2) the miscibility gap of InGaN which makes spanning the visible light spectrum difficult.\cite{Strite1992,Moustakas2001} A prospective solution to both issues is found in the II-IV-V$_2$ family of semiconductors, which are analogs to the III-Vs produced by heterovalent cation mutation (e.g. GaN becomes ZnGeN$_2$ by replacing each pair of Ga atoms with elements with one fewer and one more electron, maintaining the total charge). In general, adding elements into a compound offers more tunability than binaries, and nitride and oxynitride phase spaces are relatively unexplored.\cite{Zakutayev2016,Sun2018} Additionally, the II-IV-V$_2$ phases crystallize in either a cation ordered or disordered structure, which is predicted to tune the bandgap at constant lattice parameter, enabling control of optoelectronic properties.\cite{Martinez2017}

ZnGeN$_2$, which is analogous and lattice-matched to GaN, was first synthesized in 1970 and is theorized to have a bandgap of 3.5\,eV in its cation-ordered structure.\cite{Maunaye1970,Larson1974,Punya2011} Like the other II-IV-V$_2$ materials, which are predicted to have $\sim$1.0\,eV of bandgap tunability, the bandgap of ZnGeN$_2$ may be tunable down to $\sim$2.5\,eV by introducing cation disorder. This range spans most of the visible spectrum, which could enable lattice-matched but bandgap-tunable optoelectronics that integrate with GaN. Control of cation ordering with synthesis temperature has been demonstrated for this system, but has not yet been correlated with optical properties.\cite{Blanton2017} While high-quality crystalline and epitaxial thin films have been synthesized by sputtering, molecular beam epitaxy (MBE) and metal-organic chemical vapor deposition (MOCVD), a thorough investigation of the impact of ordering on optoelectronic properties is lacking.\cite{Narang2014,Kikkawa1999,Zhang2010,Misaki2004,Zhu1998} Photoluminescence has been demonstrated both at room temperature\cite{Zhang2010} and colder\cite{Du2008,Misaki2004,Misaki2003,Viennois2001} for cation ordered material grown by MOVPE and other vapor growth methods. Only once has photoluminescence been demonstrated for cation disordered material, in this case grown by MOCVD.\cite{Zhu1998,Zhu1999} Though bandgap tunability with ordering is an appealing goal, the optical properties of high-quality disordered material have not yet been fully explored. 

Meanwhile, the past 15 years have seen a surge in interest in the ZnGeN$_2$-ZnO alloy system for an entirely different application: photocatalytic processes.\cite{Zhang2012,Wang2008,Tessier2009,Lee2007} The majority of this experimental work has been performed by either synthesizing precursor powder of Zn$_2$GeO$_4$ or mixing Zn and Ge oxide precursors, and exposing this to ammonia while heating, creating a compound of the form (Zn$_{1+x}$Ge)(N$_2$O$_x$).\cite{Bacher1989,Zhang2012,Lee2007,Tessier2009} The structure of this alloy is reported to be wurtzite-like, with no observed peak splitting which would indicate orthorhombic ordering. Indeed, many of the reported studies of this system find that the reaction terminates in an orthorhombic (cation-ordered) ZnGeN$_2$ structure, but the orthorhombic superstructure peaks do not appear in XRD until the (nominally oxygen-free) end of the reaction.\cite{Lee2007, Zhang2012} While the bandgap of ZnO is 3.37\,eV, the alloy system with ZnGeN$_2$ exhibits bandgap bowing that lowers the optically active region to 2.0-3.0\,eV.\cite{Tessier2009, Wang2008} This is similar to the solid solution between ZnO and GaN.\cite{Xie2013} Since the lattice constants of ZnO and ZnGeN$_2$ are very similar (less than 2\% mismatch), this system offers another opportunity for bandgap tunability without significantly altering the lattice constant. No reports exist of thin films of this alloy, and the impact of cation disorder on the properties of this system has not been explored.

Here we present an epitaxial ZnGeN$_{2-x}$O$_{x}$ sample library on c-Al$_2$O$_3$ grown by combinatorial sputtering which exhibits room-temperature photoluminescence between 2.4 and 2.8\,eV. This result demonstrates that a cation-disordered ZnGeN$_{2-x}$O$_{x}$ thin film that is optically active in the green range can be grown with an inexpensive and scalable technique, indicating a path toward bandgap-tunable optoelectronic devices.

\section{\label{sec:methods}Experimental Methods}

ZnGeN$_{2-x}$O$_{x}$ thin film sample libraries were deposited by radio frequency co-sputtering onto stationary c-plane Al$_2$O$_3$ substrates with a 1$^{\circ}$ off-cut in the [100] direction heated to a set point of 750$^{\circ}$C. Metallic zinc and germanium targets angled at 45$^{\circ}$ to the substrate normal created a gradient in cation flux during synthesis, generating an array of composition conditions for each deposition. The sputtering chamber was evacuated to a base pressure of $7 \cdot 10^{-7}\,$Torr before deposition, and maintained at a working pressure of $15$\,mTorr during deposition with a gas flow of $10$\,sccm Ar and $20$\,sccm N$_2$. Oxygen was not intentionally introduced, but was incorporated from background in the sputtering chamber.

Films were first characterized using the suite of spatially resolved characterization tools available at the National Renewable Energy Laboratory, which has been used to great success for previous Zn-IV-N$_2$ work.\cite{Fioretti2015,Fioretti2018,Arca2018} Cation composition, here reported as \%Zn/(Zn+Ge), was characterized using X-ray fluorescence (XRF) with Rh L-series excitation in energy-dispersive mode using Fischer XDV-SDD software. X-ray diffraction (XRD) was performed using a Bruker D8 Discover equipped with an area detector. Transmission and reflection spectra were collected in the UV-visible (UV-Vis) spectral ranges using a custom thin film optical spectroscopy system equipped with deuterium and tungsten/halogen light sources and Si detector arrays. The spectra were then used to calculate absorption coefficient $\alpha$, using the relationship $\alpha = -$ln$[T/(1-R)]/d$, where $T$, $R$, and $d$ are transmission, reflection, and film thickness, respectively. 

Specific points on the sample library were selected for further investigation. For characterization of epitaxial films, X-ray diffraction pole figure measurements were performed using an RU200 Rigaku DMAX-A instrument with a rotating Cu K$\alpha$ anode X-ray source operating at 40\,kV/250\,mA. A Rigaku SmartLab diffractometer was used to collect reciprocal space maps.

To investigate anion composition, Rutherford backscattering spectroscopy (RBS) was performed in a 130$^{\circ}$ backscattering geometry with a 2\,MeV He+ beam energy using a model 3S-MR10 RBS system from National Electrostatics Corporation.  A 2\,MeV beam energy was used to avoid the non-Rutherford resonant scattering that occurs at 3\,MeV for nitrogen and oxygen. The 130$^{\circ}$ scattering angle was chosen to provide well separated step edges in the scattered spectrum for, in order of increasing energy, substrate oxygen, film nitrogen, film oxygen and substrate aluminum.  To obtain a spectrum with sufficient signal-to-noise ratio to observe the scattering from oxygen, RBS data was collected until the total integrated charge delivered to the sample was 400\,$\mu$C.  The film composition was determined by fitting using the RUMP analysis software\cite{Barradas2008} with the major cation ratio fixed at Zn/(Ge+Zn) = 0.51 as determined by XRF since Zn and Ge are too close in mass to be distinguished by RBS using a 2\,MeV beam energy.  

TEM micrographs were acquired with an FEI Co. Talos F200X transmission electron microscope with scanning capabilities operating at an accelerating voltage of 200\,keV. Specimens for TEM were prepared from sputtered films via in-situ focused ion beam lift-out methods\cite{Giannuzzi1999} using an FEI Co. Helios Nanolab 600i SEM/FIB DualBeam workstation. Specimens were ion milled at 2\,keV and 77\,pA to remove Ga ion beam damage and achieve a final thickness of approximately 75\,nm. Structural characterization was conducted by acquiring selected area electron diffraction (SAED) patterns on an FEI Co. Ceta 16M pixel CMOS camera at a camera length of 660\,mm. The Al$_2$O$_3$ substrate was used to calibrate the camera constant, allowing SAED reflections to be accurately measured and indexed. Chemical mapping was performed in the TEM using the Super-X energy-dispersive X-ray spectroscopy (EDS) system.

Photoluminescence was acquired using a Newport MS260i imaging spectrometer, an InstaSpec X array detector with a 300/500 grating and an exposure time of 5 seconds. The PL was measured at room temperature with a 325\,nm laser line at 4.5\,mW. The CCD UV edge was at 400\,nm and the long pass filter used was at 365\,nm. Tapping mode atomic force microscopy (AFM) was performed on an Asylum Research system using a tip with a 150\,N/m force constant to characterize surface morphology.

\section{\label{sec:results}Results}

In this study, the substrate temperature and gun powers were fixed, resulting in a sample library with a linear composition gradient. This library consisted of 9 unique cation compositions from Zn/(Zn+Ge) = 0.47 to 0.51. After all spatially resolved characterization was performed, the library was cleaved into smaller pieces for single-point measurements (such as XRD pole figures, RBS and TEM).

\subsection{\label{sec:diffraction}X-ray Diffraction}

\begin{figure}[t]%
\includegraphics*[width=1.0\linewidth]{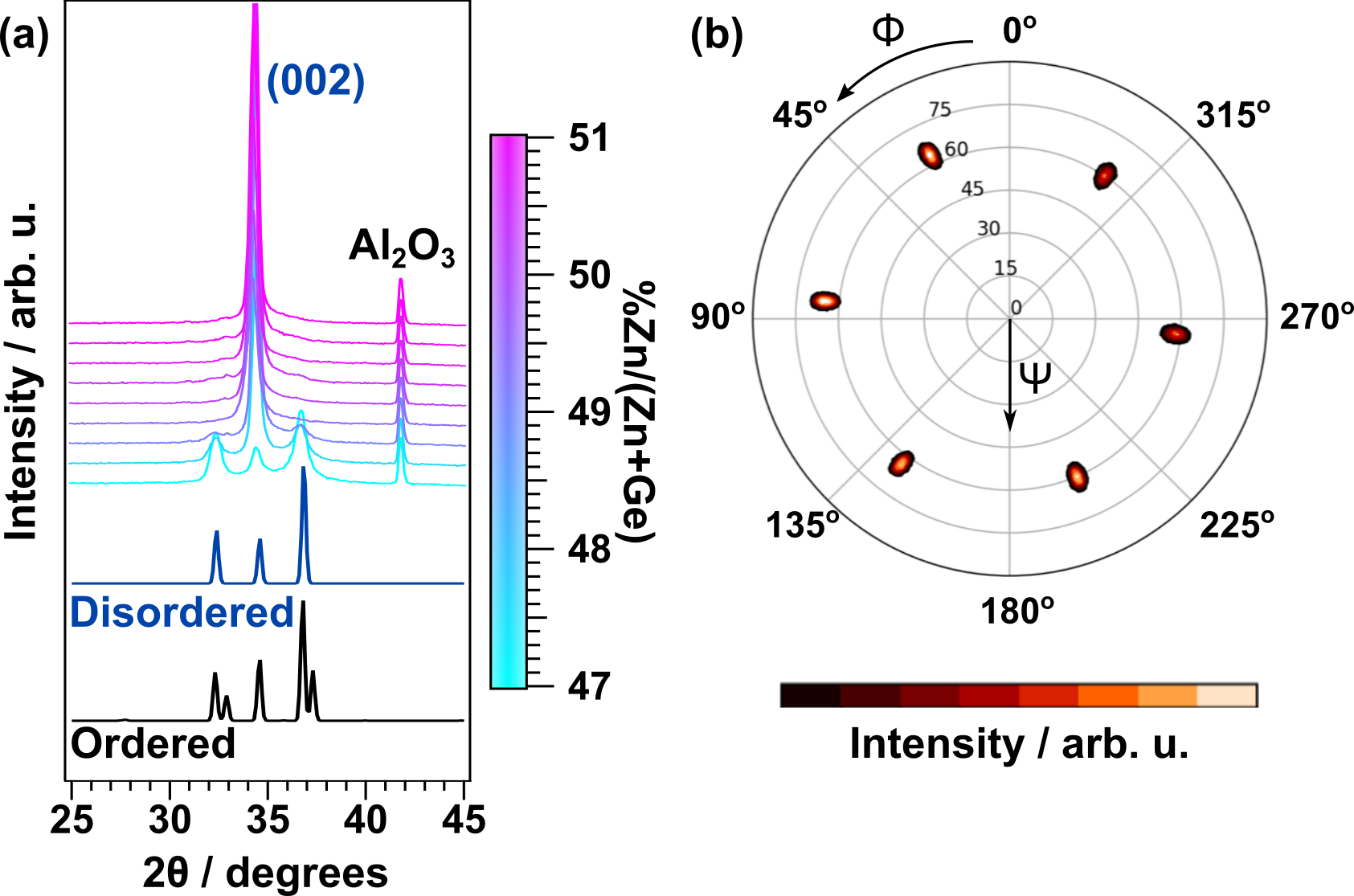}
\caption{%
(a) X-ray diffraction patterns as a function of cation composition (color axis). At lower Zn concentrations, three peaks indicating wurtzite ZnGeN$_{2-x}$O$_{x}$ are present, in addition to the Al$_2$O$_3$ peak. At stoichiometric and higher Zn concentrations, only the (002) wurtzite peak is present. (b) X-ray pole figures, which investigate in-plane orientation of a film, reveal 6-fold symmetry of the (101) peak indicating epitaxial alignment of the ZnGeN$_{2-x}$O$_{x}$ film with the Al$_2$O$_3$ substrate. The film is 30$^{\circ}$ rotated from the substrate (Fig. \ref{suppPoleFigs}), which is the same epitaxial relationship as GaN on c-Al$_2$O$_3$.\cite{Itoh1985}
  }
\label{xrd}
\end{figure}

X-ray diffraction patterns are shown as a function of cation composition in Fig.~\ref{xrd}a. At lower ($\sim$47\%) Zn concentrations, three wurtzite peaks are present. No superstructure peaks or peak splitting are visible, indicating cation-disordered material has been synthesized.\cite{Blanton2017} The Al$_2$O$_3$ (001) substrate peak is also present, and no additional peaks indicating secondary phases are observed. As the Zn concentration increases, the wurtzite (002) peak increases in intensity and all other peaks are no longer observed, indicating a high degree of c-plane texturing in the film.

While the polycrystalline side of this library is easily identifiable as wurtzite from X-ray diffraction, the strongly textured side reveals a film peak position ranging from 34.3 to 34.5 degrees. This is slightly smaller than the theoretical values for the (002) peaks for the wurtzite and orthorhombic ZnGeN$_2$ structures, which are 34.56 degrees and 34.51 degrees respectively.\cite{Larson1974,Maunaye1970} Though slightly shifted, this peak position is consistent with reported values (002) peak of ZnGeN$_2$, ZnO, and ZnGeN$_{2-x}$O$_{x}$, since these structures are all wurtzite and have a very similar lattice constant. In order to further investigate structure, X-ray pole figures were performed on the points where XRD indicated strong texturing. Peaks are present at the 2$\theta$ and $\Psi$ positions that correspond to the wurtzite (101) peak, indicating that the film is indeed crystallized in the wurtzite structure. A representative scan of the (101) peak is shown in Fig.~\ref{xrd}b. Six-fold symmetry indicates epitaxial alignment of the film with the Al$_2$O$_3$ substrate. The pole figures reveal a 30$^{\circ}$ in-plane rotation between the film and the substrate (Fig.~\ref{suppPoleFigs}), the same epitaxial relationship as between GaN and Al$_2$O$_3$.\cite{Itoh1985}  

\begin{figure}[!t]
\includegraphics[width=0.95\linewidth]{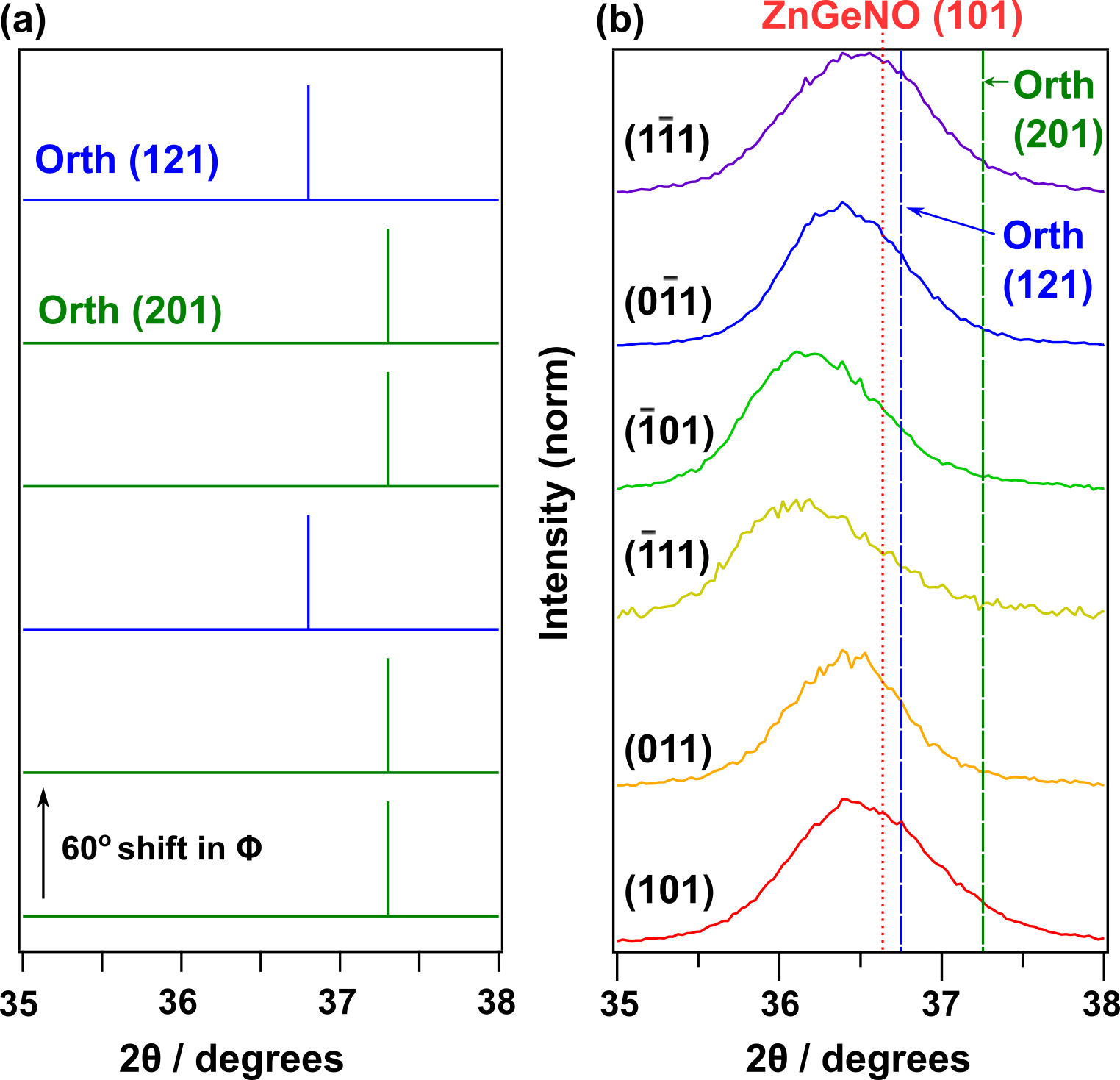}
\caption{(a) Simulated peak positions in 2$\theta$ for ordered (orthorhombic) ZnGeN$_2$, each with a rotation of 60$^{\circ}$ in $\phi$. The (201) family of reflections (green) exhibit the same position in 2$\theta$, while the two peaks in the (121) family (blue) exhibit a distinct higher position in 2$\theta$. (b) Line scans in 2$\theta$ from each RSM with each color at a rotation of $\sim$60$^{\circ}$ in $\phi$. The traces are indexed assuming a wurtzite structure. The slight shifts in 2$\theta$ are likely due to a small instrument misalignment, and no trend is seen indicating orthorhombic distortion when compared to (a). The red reference line is for the (101) peak of Zn$_{1.231}$Ge$_{0.689}$O$_{0.782}$N$_{1.218}$ reported in Ref.~\citenum{Bacher1989}, while the blue and green reference lines are the orthorhombic ZnGeN$_2$ (121) and (201) positions shown in (a).}
\label{RSM}
\end{figure}

Reciprocal space maps (RSMs) reveal full-width at half-max (FWHM) values of 120 arcminutes for the (002) peak and 150 arcminutes for the (101) peak. In comparison, Daigo et. al. measured FWHM values of 100-140 arcminutes for a GaN film grown by sputter epitaxy on Si.\cite{Daigo2005} Additional RSMs were performed to determine whether any cation ordering was occurring in the epitaxial sample. Orthorhombic distortion due to ordering would cause systematic peak shifting from the ideal wurtzite structure as a function of $\phi$, as simulated in Fig.~\ref{RSM}a. The (101) peak of the epitaxial film is observed to shift slightly to 2$\theta$ values lower than the ideal ZnGeN$_2$ structures, which is consistent with the incorporation of oxygen, as shown in Fig.~\ref{RSM}b. The (101) peak displays small shifts with $\phi$, but no systematic shifts that would indicate cation ordering (Fig.~\ref{RSM}b). The observed peak shifts follow a sinusoid that is coincident with the measured alignment offset of the instrument. It is important to note that across the X-ray scan area, the superposition of cation-ordered orthorhombic grains with different in-plane orientations could yield a superposition of the orthorhombic (121) and (201) peaks shown in Fig.~\ref{RSM}a. It is possible that the low signal and broad XRD peaks of this sputtered film are masking this type of spatially inhomogenous ordering in Fig.~\ref{RSM}b. However, the width of the asymmetric scan is the same order of magnitude as the symmetric scan rocking curve, which is not sensitive to ordering, suggesting that the broadening is due to sputtered material quality and not orthorhombic distortion. Additionally, considering that the ordered orthorhombic structure of ZnGeN$_2$ has never been synthesized below 850$^{\circ}$C,\cite{Blanton2017} we hypothesize that this film is disordered.

\subsection{\label{sec:composition}Composition}

\begin{figure}[t]%
\includegraphics*[width=1.0\linewidth]{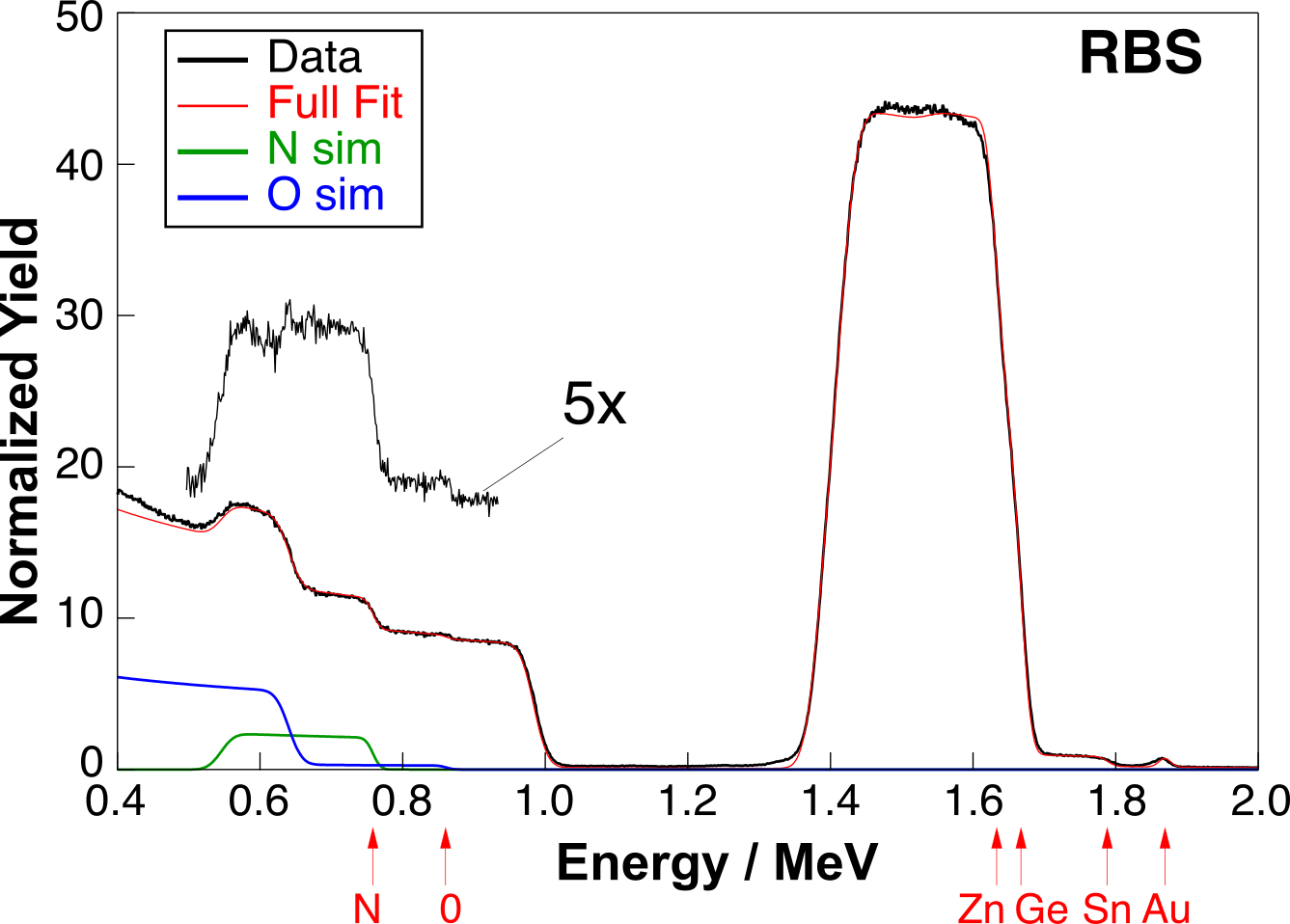}
\caption{Rutherford backscattering (RBS) analysis of one ZnGeN$_{2-x}$O$_{x}$ film on Al$_2$O$_3$ substrate with thin Au coating to avoid sample charging.  Black:  measured spectrum; Red: full fit to three layer model (Au/Zn-Ge-Sn-N-O/Al$_2$O$_3$); Green: N contribution to fit; Blue; O contribution to fit.  Vertical red arrows indicate the front surface scattering energy for elements as labelled.  The inset spectrum section shows the measured oxygen and nitrogen region with the modeled substrate signal subtracted expanded by 5X and offset for clarity.}
\label{RBS}
\end{figure}

Specific points on the sample library were selected for Rutherford backscattering (RBS) analysis to quantify oxygen and nitrogen content (Fig.~\ref{RBS}). A thin gold coating was used to eliminate sample charging issues. Analysis was performed with different starting compositions in order to determine a standard error for the model. The Zn and Ge compositions were set by X-ray fluorescence to be 1.02 and 0.98 formula unit, respectively. 
The film was also found to contain 1 at.\,\% of a heavy contaminant, here assumed to be Sn due to previous sputtering with this element in the synthesis chamber. 
This yields a compound composition of Zn$_{1.02}$Ge$_{0.98}$N$_{1.91}$O$_{0.18}$, with an anion-to-cation ratio of 1.05 and an oxygen composition of 8.8\% O/(O+N). This is on the lower end of the reported oxygen content for ZnGeN$_{2-x}$O$_x$ material (discussed further in Section~\ref{sec:discussion}). Quantified EDS analysis shows uniform composition spatially throughout the investigated film, and confirms the composition numbers from RBS to an order of magnitude. EDS shows some carbon contamination as well, which mainly occurs at the substrate-film interface. It is promising that despite the Sn and C contaminants, the film is optically active and of relatively high crystalline quality.

\subsection{\label{sec:TEM}Microscopy}

\begin{figure}[t!]%
\includegraphics*[width=0.95\linewidth]{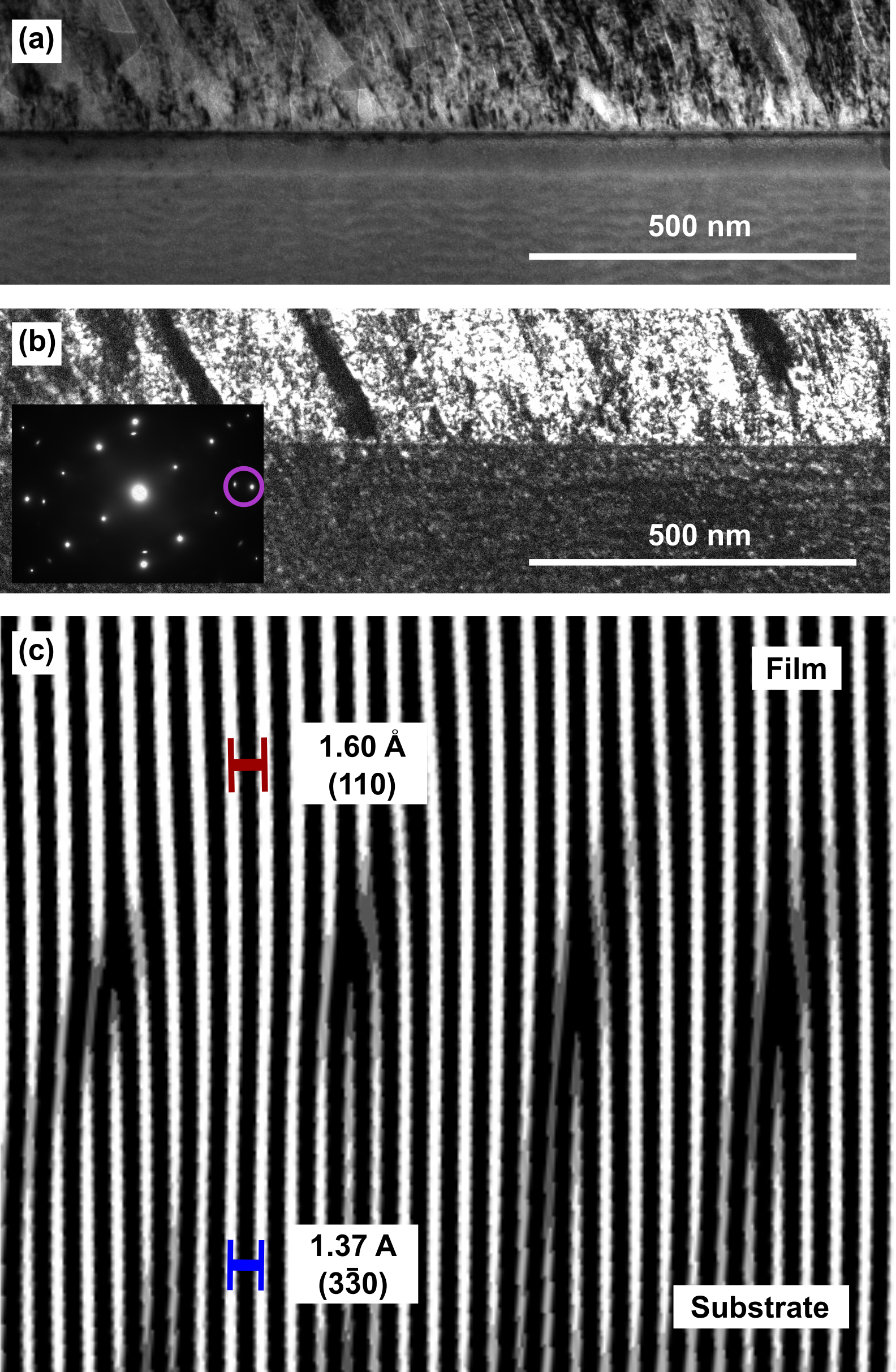}
\caption{%
 (a) Bright-field TEM reveals a uniform, columnar film approximately 200\,nm thick. (b) Dark-field TEM acquired using the film (110) and the substrate (3$\overline{3}$0) peaks (circled in selected area electron diffraction, inset). The film is non-uniform and shows a few dark regions indicating non-epitaxial grains. However, the majority of the film is aligned with the substrate and the same orientation. The substrate appears darker than the film due to the lower intensity of the substrate peak. (c) Fourier filtered high-resolution TEM micrograph showing a semi-coherent substrate-film interface with periodic dislocations approximately every 7-8 lattice planes to relieve strain due to large lattice mismatch between ZnGeN$_{2-x}$O$_{x}$ and Al$_2$O$_3$. (The original lattice image is shown in Fig.~\ref{suppTEM}.)
  }
\label{TEM}
\end{figure}

Figure~\ref{TEM}a-b shows TEM bright-field and dark-field micrographs of the epitaxial ZnGeN$_{2-x}$O$_{x}$ film on Al$_2$O$_3$ substrate. Selected area electron diffraction in the inset of (b) shows reflections corresponding to the [110] zone axis of the Al$_2$O$_3$ substrate and the [1$\overline{1}$0] zone axis of the ZnGeN$_{2-x}$O$_{x}$ film; the purple circle indicates the (110) film and (3$\overline{3}$0) substrate reflections used for dark-field imaging. As expected for a combinatorial sample grown by sputtering, the film is non-uniform and shows some dark regions indicating non-epitaxial grains. However, the majority of the film is aligned with the substrate and a consistent crystallographic orientation. Fig.~\ref{TEM}c shows a Fourier filtered high-resolution TEM micrograph of the substrate-film interface. Peaks in the Fourier transform of the as-acquired micrograph corresponding to the Al$_2$O$_3$($3\overline{3}$0) and ZnGeN$_{2-x}$O$_{x}$(110) lattice planes were selectively filtered out to better view their crystallographic relationship. A semi-coherent interface is observed exhibiting periodic dislocations every 7-8 lattice planes which relieve strain due to lattice mismatch between the film and substrate. 

The X-ray diffraction results from Section~\ref{sec:diffraction} and the SAED results allow us to calculate the epitaxial alignment between substrate and film. Since only the (002) ZnGeN$_{2-x}$O$_{x}$ peak appears by X-ray diffraction, this suggests that the c-planes are aligned between the film and the substrate. This is confirmed by SAED, which demonstrates that ZnGeN$_{2-x}$O$_{x}$ (002) is parallel to Al$_2$O$_3$ (006). With regard to in-plane alignment, the X-ray pole figures reveal a 30$^{\circ}$ rotation between the film and the substrate. This is also confirmed by TEM: Fig.~\ref{TEM}c demonstrates alignment between the ZnGeN$_{2-x}$O$_{x}$ (110) planes and the Al$_2$O$_3$ ($3\overline{3}$0) planes, which corresponds to a 30$^{\circ}$ rotation. This is the same relationship as reported previously for ZnGeN$_2$ on c-Al$_2$O$_3$, and for GaN on c-Al$_2$O$_3$.\cite{Zhu1998,Itoh1985} Based on these in-plane and out-of-plane relationships, the calculated in-plane lattice mismatch is 13.3\%, which is reflected in the SAED and the high-resolution TEM (Fig.~\ref{TEM}b-c). 

Interestingly, the reciprocal space maps, pole figures and SAED all reveal a $\sim$3.6$^{\circ}$ tilt between the film and the substrate. While Nagai tilt has been observed in GaN due to substrate off-cut, the tilt we observe is rotated 90$^{\circ}$ from the angle of the off-cut, inconsistent with the Nagai model.\cite{Huang2005} The origin of this tilt is unknown and warrants further exploration.

Finally, AFM was performed at multiple points on the film to investigate surface morphology and growth microstructure (Fig.~\ref{suppAFM}). Scans revealed a dense, round microstructure with features ranging 75-250\,nm diameter, with RMS roughness of 5-6\,nm. Four-point electrical measurements were also performed, and the film was found to be insulating. This is expected considering the insulating properties of ZnGeN$_2$ predicted by DFT.\cite{Skachkov2016}

\subsection{\label{sec:optical}Optical Properties}

\begin{figure}[t]%
\includegraphics*[width=\linewidth]{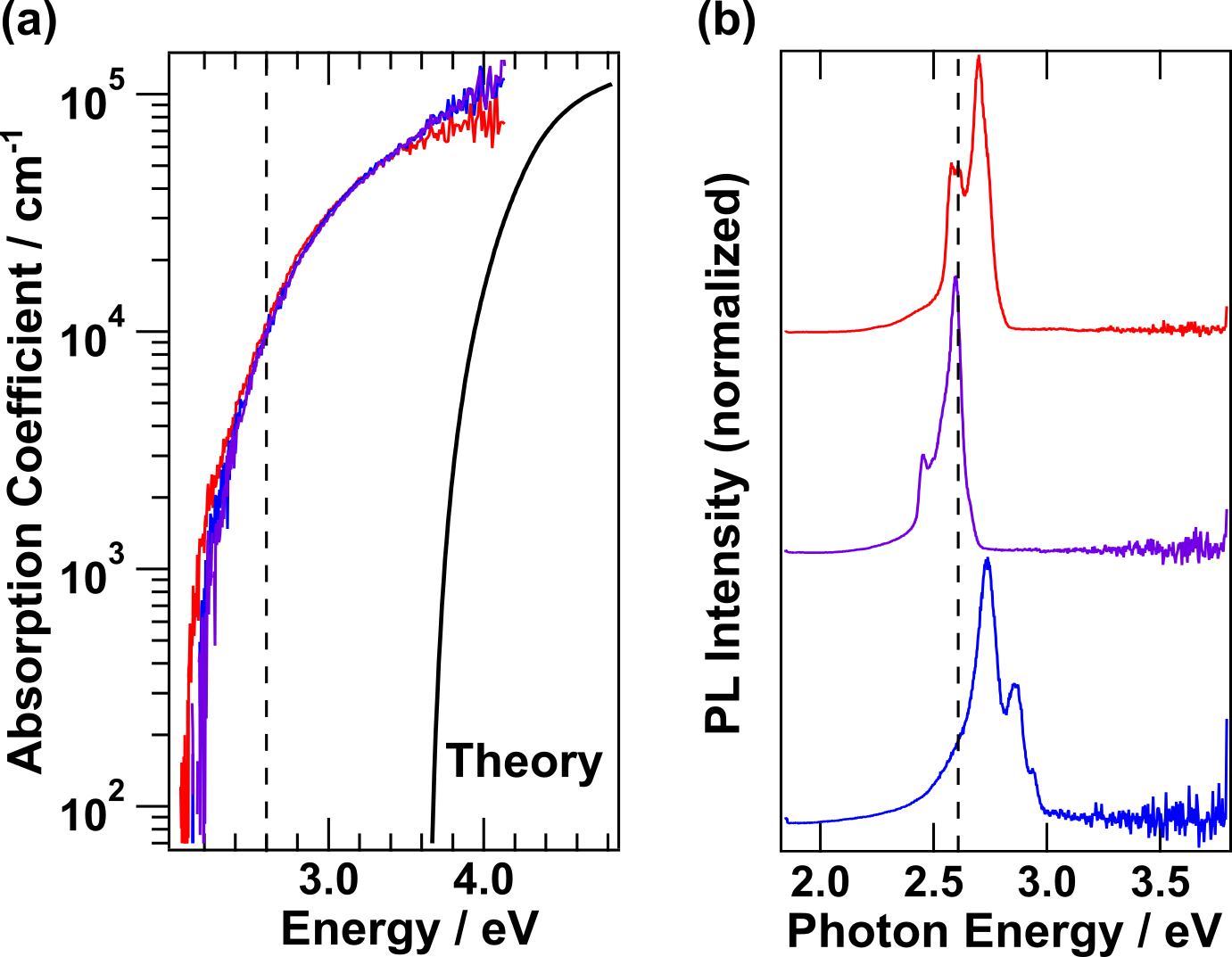}
\caption{%
 (a) UV-Vis Spectroscopy reveals an absorption onset at approximately 2.6\,eV for all samples. This is $\sim$1.0\,eV lower than the theoretical absorption onset for ordered (orthorhombic) ZnGeN$_2$, shown as a black curve, consistent with bandgap tuning due to oxygen incorporation. The dashed line at 2.6\,eV represents the energy at an absorption coefficient of 10$^4$\,cm$^{-1}$.  Colors are added to distinguish individual samples used in both UV-vis and PL. DFT calculations reproduced from Ref. \cite{Martinez2017}. (b) Room-temperature photoluminescence of ZnGeN$_{2-x}$O$_{x}$ films reveal peaks from 2.4 to 2.8\,eV. Different points on the sample yield a slightly different set of peak positions and intensities. No trend is observed with cation composition variation. These values align with the observed 10$^4$ absorption coefficient value from the UV-vis spectroscopy (dashed line). 
 }
\label{PL_uvVis}
\end{figure}

In order to determine the feasibility of incorporation of ZnGeN$_{2-x}$O$_{x}$ into optoelectronic devices, optical properties of these films were studied using UV-visible spectroscopy and room-temperature photoluminescence. Figure~\ref{PL_uvVis}a shows the absorption coefficient from UV-vis spectroscopy as a function of energy, with different colors representing different points on the sample library. The absorption onset is $\sim$1.0\,eV lower than the theoretical absorption onset for ordered ZnGeN$_2$ (shown as a solid black line), but remains a similar shape to theory, without band tailing that would indicate a high defect concentration. The energy value at an absorption coefficient of 10$^4$/cm is 2.6\,eV, shown as a dashed line. Little change is observed with cation composition. Normalized room-temperature photoluminescence for different points on the sample library is shown in Fig.~\ref{PL_uvVis}b. Though the PL signal was spatially varying throughout each film, bright spots were identified, revealing peaks from 2.4 to 2.8\,eV. The peak variation does not correlate with changes in cation composition. These values are consistent with the UV-vis optical absorption onset of 2.6\,eV (dashed line). Multiple varying-energy PL peaks is characteristic of a material exhibiting potential fluctuations near the band edge, which is expected in materials with structural disorder.\cite{Scragg2016,Fioretti2018}

Due to the alignment of our PL results with the measured absorption data, we hypothesize that our PL emission is due to band-like recombination in the ZnGeN$_{2-x}$O$_{x}$ thin film. Though we could be observing ``yellow band'' defect states, to which other groups have attributed PL emission between 2 and 3\,eV for cation-ordered ZnGeN$_2$, the traditional yellow band is $\sim$0.5\,eV wide.\cite{Du2008} The PL peaks reported here exhibit an average FWHM value of 0.16\,eV when each spectrum is fit with one Gaussian envelope. When each individual peak is fit, the average FWHM value is 0.08\,eV. These widths are narrow compared to traditional defect band luminescence, and since the PL energies match the sharp UV-vis absorption onset and the high-energy tails follow a Maxwell-Boltzman distribution, this suggests that the optical signal originates from band-like luminescence.

\subsection{\label{sec:discussion}Discussion}

\begin{figure}[t]%
\includegraphics*[width=0.95\linewidth]{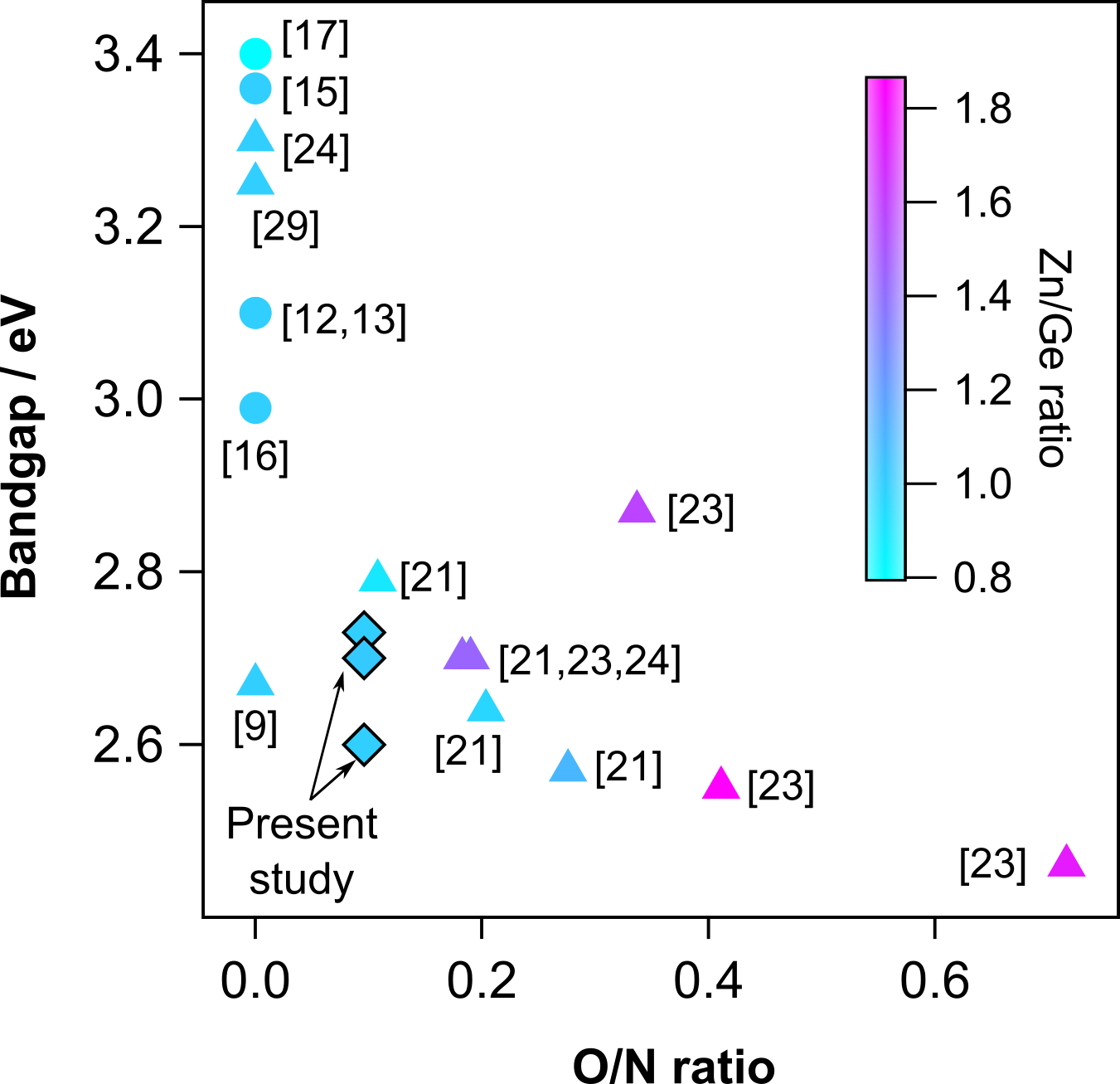}
\caption{Plot of reported bandgap versus O/N ratio from the literature, with triangular traces denoting a reaction process using oxide precursors. Circular traces denote a reaction process from metallic and/or nitride precursors. Diamonds represent the PL peak positions reported in this work (Fig.~\ref{PL_uvVis}b). Zn/Ge ratio is plotted on the color axis.}
\label{bandgap}
\end{figure}

To place our results in historical context, data of reported bandgap versus O/N ratio was collected from the literature and is shown in Fig.~\ref{bandgap}. Triangular traces are plotted for material grown using a reaction process starting from oxygen-containing precursors, while circular traces denote a reaction from metallic and/or nitride precursors. Diamond traces represent values reported in this study. A clear decrease in bandgap is observed with increasing oxygen content that follows the energy bandgap bowing trend. The cation composition (color axis in Fig.~\ref{bandgap}) follows the expected trend as well, since an alloy between ZnGeN$_2$ and ZnO implies a trend of increasing Zn content with increasing O content. Many of the data points that report pure ZnGeN$_2$ material (corresponding to an O/N ratio of 0) do not report characterization of oxygen, suggesting that some unintentional oxygen content may be decreasing the bandgap and contributing to the large bandgap spread at O/N$=$0.

Though all reported ZnGeN$_{2-x}$O$_{x}$ material exhibits the disordered wurtzite structure, one could conceive of a cation-ordered ZnGeN$_{2-x}$O$_{x}$ structure of the same orthorhombic superstructure as the ground state of ZnGeN$_2$. This raises a few questions about the ZnGeN$_{2-x}$O$_{x}$ system: Could cation ordering be used to tune the optical properties of ZnGeN$_{2-x}$O$_{x}$ at a fixed composition and lattice parameter? On the other hand, could a small amount of oxygen be used to trap ZnGeN$_2$ in a cation-disordered state? Control of cation ordering has been a sought-after research goal in the ZnGeN$_2$ system for optoelectronic researchers, but the presence of oxygen (intentional or not) may help or hinder this prospect. From a physical vapor deposition perspective, oxygen often incorporates unintentionally into thin films due to background contamination in vacuum systems, causing deleterious defects in material systems that require high purity. This work demonstrates that harnessing oxygen incorporation could provide an additional knob with which to tune properties of the ZnGeN$_{2-x}$O$_x$ system. Since this alloy exhibits only small changes in lattice parameter, this provides a tunable bandgap material from 2.4\,eV to 3.4\,eV that is lattice matched to GaN.

It is also important to note the diversity of characterization methods used to collect the bandgaps shown in this plot; the studies on ZnGeN$_2$ thin films typically report bandgaps from PL\cite{Du2008,Misaki2004,Viennois2001} or Tauc plots from UV-Vis data,\cite{Zhu1999,Narang2014} while most ZnGeN$_{2-x}$O$_{x}$ studies report Kubelka-Munk absorbance curves.\cite{Zhang2012,Lee2007,Tessier2009} We have plotted PL peak positions (2.60, 2.70, and 2.73\,eV). None of these optical characterization methods are perfect, and it is evident that more data is necessary to deconvolve these results. Additionally, while Fig.~\ref{bandgap} shows the impact of Zn/Ge and O/N on the bandgap, some of the variation in the data may be due to order parameter, which has not been taken into account. Further optical and structural characterization is key to deconvolve the impacts of cation composition, anion composition, and disorder on optical properties of ZnGeN$_{2-x}$O$_x$.

\section{\label{sec:conclusion}Conclusion}

In this work, we have reported an epitaxial ZnGeN$_{2-x}$O$_{x}$ sample library grown on c-plane Al$_2$O$_3$ by sputtering. X-ray diffraction and TEM confirm epitaxial alignment with the substrate despite 13.3\% in-plane lattice mismatch. The sputter-deposited samples are optically active with room-temperature PL peaks that are narrow compared to traditional defect band luminescence. These peaks align with a sharp optical absorption onset at 2.6\,eV, and we hypothesize this optical signal originates from band-related luminescence in ZnGeN$_{2-x}$O$_{x}$. To the authors' knowledge, this result is the first reported ZnGeN$_{2-x}$O$_{x}$ thin film, and sets the stage for low-cost light-emitting devices which integrate with GaN based on the ZnGeN$_2$-ZnO material system.

\section{\label{sec:ack}Acknowledgments}

This work was authored in part by Alliance for Sustainable Energy, LLC, the manager and operator of the National Renewable Energy Laboratory for the U.S. Department of Energy (DOE) under Contract No. DE-AC36-08GO28308. This work was supported by the U.S. Department of Energy, Office of Science, Basic Energy Sciences, Materials Sciences and Engineering Division. The views expressed in the article do not necessarily represent the views of the DOE or the U.S. Government. The U.S. Government retains and the publisher, by accepting the article for publication, acknowledges that the U.S. Government retains a nonexclusive, paid-up, irrevocable, worldwide license to publish or reproduce the published form of this work, or allow others to do so, for U.S. Government purposes.

\bibliographystyle{apsrev4-1}

\newcommand{\beginsupplement}{%
        \setcounter{table}{0}
        \renewcommand{\thetable}{S\arabic{table}}%
        \setcounter{figure}{0}
        \renewcommand{\thefigure}{S\arabic{figure}}%
     }
     
\beginsupplement

\newpage

\begin{figure*}[!b]
\centering
\includegraphics[width=13cm]{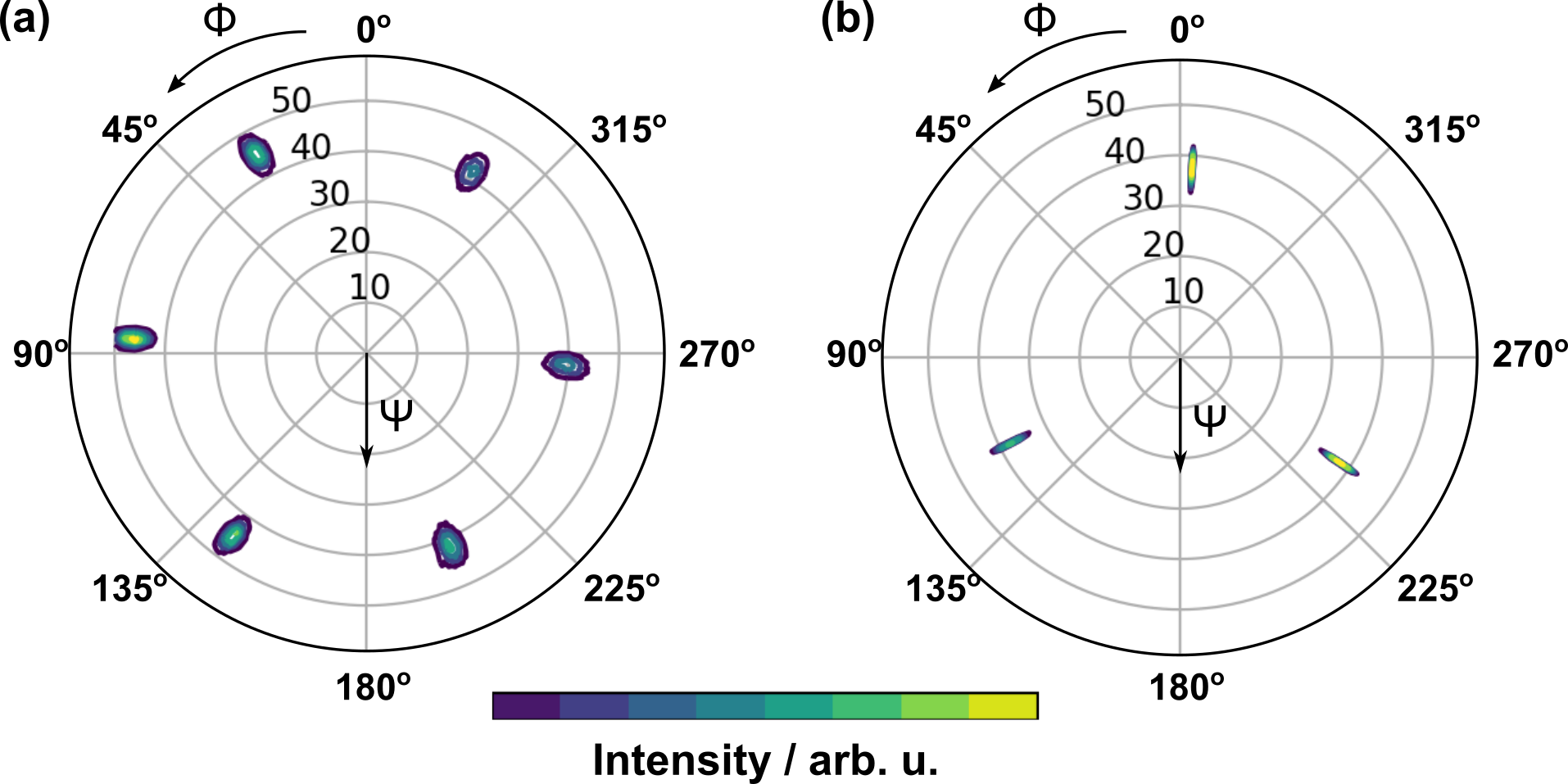}
\caption{(a) X-ray pole figure of the (102) ZnGeN$_{2-x}$O$_x$ peak. (b) Pole figure of the (104) Al$_2$O$_3$ peak, demonstrating the 30$^{\circ}$ rotation between the film and substrate. Both scans were taken in the same session without realignment of the film.}
\label{suppPoleFigs}
\end{figure*}

\begin{figure*}[!b]
\centering
\includegraphics[width=17cm]{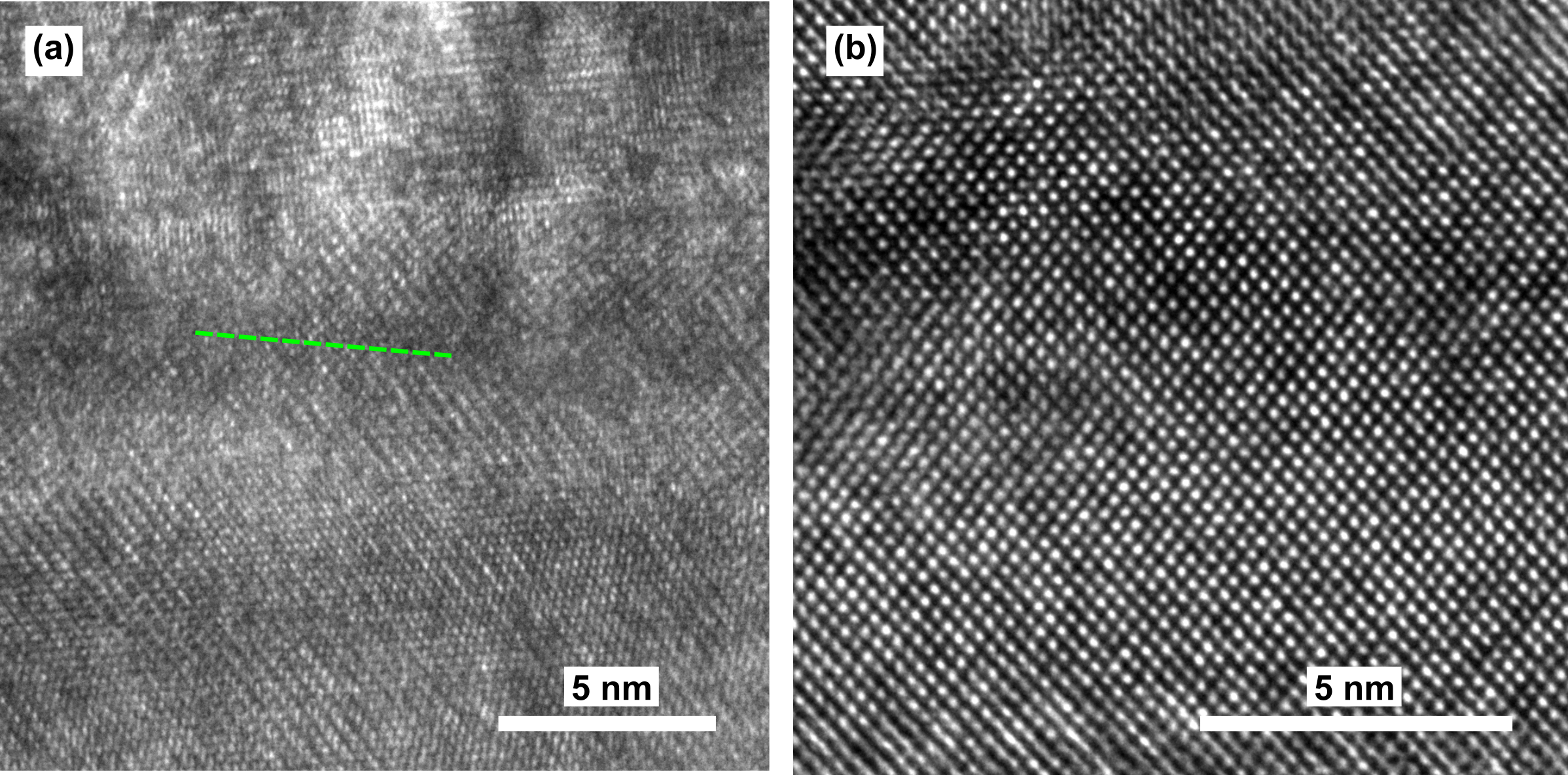}
\caption{(a) HRTEM image of the interface between the substrate and the film. The interface shown in the Fourier filtered image in Fig.~\ref{TEM} is marked with a dashed green line. (b) HRTEM image of the film near the surface.}
\label{suppTEM}
\end{figure*}

\begin{figure*}[!b]
\centering
\includegraphics[width=8cm]{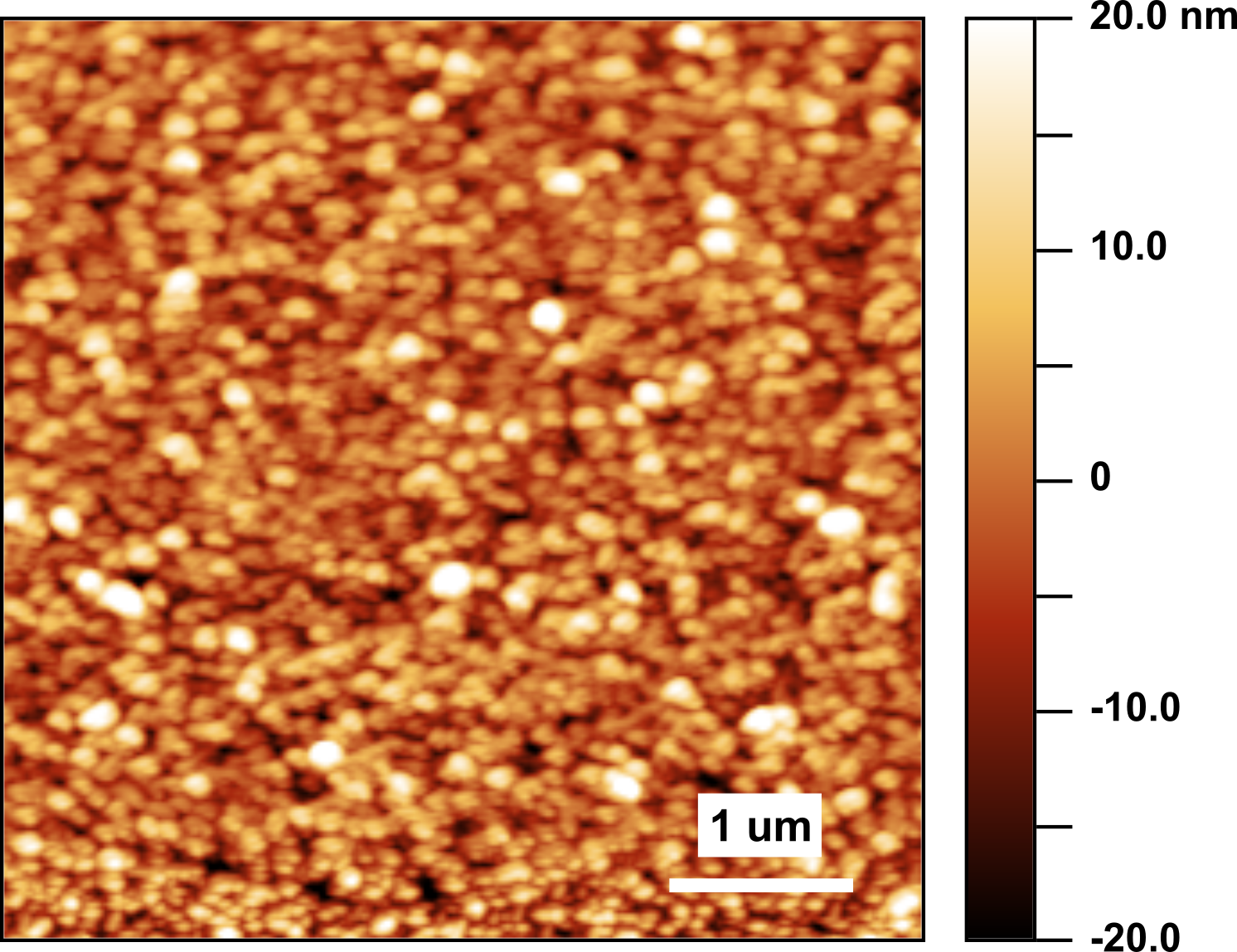}
\caption{Atomic force microscopy reveals dense, round microstructure with features from 75-250\,nm in diameter. The RMS roughness ranges from 5-6\,nm across multiple scans.}
\label{suppAFM}
\end{figure*}

\end{document}